\documentclass[twocolumn]{aastex701}
\usepackage{graphicx}
\usepackage{amsmath}
\usepackage{booktabs}
\usepackage{multirow}
\usepackage{float}
\usepackage[caption=false]{subfig}
\usepackage{xcolor}
\usepackage{natbib}
\usepackage{tikz}
\usetikzlibrary{arrows.meta,positioning}

\usepackage{soul}

\shorttitle{SHAPE: Flare Forecasting}
\shortauthors{S.Kadakol et al.}

\begin{document}

\title{SHAPE: Solar Heliospherical Analysis and Prediction of Eruptions\\
        A machine learning approach to solar flare forecasting}

\correspondingauthor{S. Kadakol}
\email{suvan.kadakol@mitwpu.edu.in}

\author[0009-0009-2269-9469]{S. Kadakol}
\email{suvan.kadakol@mitwpu.edu.in}
\affiliation{Department of Physics, MIT World Peace University, Pune 411038, India}

\author[0009-0002-5823-9930]{P. Deshmukh}
\email{param.deshmukh@mitwpu.edu.in}
\affiliation{Department of Physics, MIT World Peace University, Pune 411038, India}

\author[0000-0001-5208-8989]{A. B. Bendre}
\email{abhijit.bendre@mitwpu.edu.in}
\affiliation{Department of Physics, MIT World Peace University, Pune 411038, India}

\begin{abstract}
Solar flares are the most energetically impulsive events in the solar system, and their consequences for near-Earth infrastructure are well established. Reliable operational forecasting, however, particularly of the rarest and most destructive M- and X-class events, remains an open problem. We present Solar Heliospherical Analysis and Prediction of Eruptions (SHAPE), a two-phase deep-learning pipeline for solar flare forecasting that deliberately bypasses sunspot proxies in favour of direct soft X-ray observations from the X-ray Sensor (XRS) on the Geostationary Operational Environmental Satellites (GOES). Phase 1 forecasts monthly flare counts, aggregated to annual totals, across whole solar cycles with a hybrid model: a cycle-shape curve anchored to published solar-cycle minima fixes the timing and mean profile of the cycle, and a compact gated-recurrent-unit network with Monte Carlo dropout models the residual variability around it. We train on the science-quality NOAA GOES Flare Report for 1995--2020, which already incorporates retrospective calibration and outage fixes, and validate on the unseen years 2021--2025. Phase 1 achieves a mean absolute error of 204 flares yr$^{-1}$ over this period, although it underestimates the 2024 solar maximum. Phase 2 forecasts the next day's strongest flare class (C, M or X) from a shared-trunk, dual-head LSTM operating on a 12-day rolling window of five lagged and rolling-maximum intensity features, with an optional deterministic blend against an active-region physics score and a threshold-based decision rule tuned to favour recall of M-or-above (M+) events. On a 100-day chronological backtest, the classifier attains a true skill statistic of 0.351 and a Heidke skill score of 0.335 for M+ events, a Brier skill score of 0.163 relative to climatology, and well-calibrated probabilities (expected calibration error 0.023). A simple day-to-day persistence baseline is, however, competitive with the neural classifier on both skill scores, underscoring how much of the short-term predictability in flare class is autocorrelation rather than learned structure. SHAPE demonstrates that competitive solar flare forecasting is achievable using architectures compact enough to retrain in minutes on standard hardware, without any sunspot proxy inputs.
\end{abstract}


\section{Introduction}
\label{sec:intro}

Solar flares are characterised by the impulsive release of magnetic free energy stored in coronal active regions, which heats coronal plasma, accelerates particles and enhances the electromagnetic emission from radio waves to hard X-rays \citep{shibata2011, benz2017, fletcher2011}. The enhanced X-ray and extreme-ultraviolet radiation reaches the Earth after about eight minutes, whereas energetic particles from the associated eruptions arrive after tens of minutes to hours \citep{desai2016}. The operational impact of these events on near-Earth infrastructure is well documented \citep{cannon2013, eastwood2017} and proceeds through several distinct channels. The flare radiation itself enhances ionisation in the dayside ionosphere, causing high-frequency radio blackouts \citep{davies1990}, and the intense radio bursts that can accompany flares degrade the tracking of Global Positioning System (GPS) signals \citep{cerruti2008}. The coronal mass ejections (CMEs) associated with large flares, rather than the flares themselves, are the principal drivers of intense geomagnetic storms \citep{gosling1993, kilpua2017}, which induce potentially damaging currents in high-voltage transmission networks \citep{pulkkinen2017}. Solar energetic particle events further raise the radiation dose received at aviation altitudes \citep{meier2014}.

The 1859 Carrington event, in which a white-light flare observed by \citet{carrington1859} preceded an extreme geomagnetic storm \citep{tsurutani2003}, remains the benchmark extreme case. A comparable event under present-day technological dependence has been estimated to cost of order one to two trillion US dollars in the first year alone for the United States \citep{nrc2008}, with similar estimates for the North American electricity grid \citep{lloyds2013}. The October--November 2003 ``Halloween'' storms produced spacecraft anomalies and outages across many missions \citep{webb2004}. Solar Cycle 25 has, moreover, already exceeded the consensus forecast: its smoothed sunspot number peaked at approximately 161 in October 2024 \citep{sidc2025}, about 40\% above the value of 115 predicted by the 2019 NOAA/NASA Solar Cycle 25 Prediction Panel \citep{swpc_panel2019}. Such misses underscore persistent limitations of current forecasting capability \citep{petrovay2020}.

Modern infrastructure increasingly relies on precise satellite positioning and timing, for example in precision agriculture \citep{schimmelpfennig2016}, in the time-stamping of high-frequency financial trading, where European regulation requires traceability to Coordinated Universal Time within 100~$\mu$s \citep{eu2017clocks}, and in navigation systems aided by the Global Navigation Satellite System (GNSS) \citep{cannon2013}. Each of these is sensitive, directly or indirectly, to space weather \citep{eastwood2017}. Flare forecasting has consequently shifted from a specialised concern of the radio and space-physics communities to an element of infrastructure resilience planning \citep{schrijver2015, cannon2013}.

Existing flare-forecasting approaches fall broadly into two families. Event-based methods, with lead times of hours to days, estimate the probability that an active region will produce a flare above a given class within a forecast window. They use magnetic classifications of active regions such as the McIntosh scheme \citep{mcintosh1990}, Poisson or Bayesian statistics of past flaring \citep{gallagher2002, wheatland2004}, or machine-learning classifiers trained on photospheric magnetic-field parameters, such as the Space-weather HMI Active Region Patches (SHARP; \citealt{bobra2014}) derived from the Helioseismic and Magnetic Imager (HMI; \citealt{schou2012}) on the Solar Dynamics Observatory (SDO; \citealt{pesnell2012sdo}) \citep{bobra2015, nishizuka2017, florios2018, liu2017}. Systematic comparisons show that the skill of these methods is modest and strongly dependent on the flare-class threshold \citep{barnes2016, leka2019}. Cycle-scale methods, with lead times of years, forecast the smoothed sunspot number using precursor, extrapolation, dynamo-based and, increasingly, deep-learning techniques \citep{pesnell2012cycle, petrovay2020, pala2019, benson2020}; the analogous forecasting of flare activity at cycle scale has only recently been attempted \citep{cao2025}.

Both families rely on sunspot- or active-region-derived proxies rather than on the X-ray emission itself, which limits them in four ways. First, the proxies describe the magnetic or morphological state of the photosphere, whereas the quantity of interest, soft X-ray emission from flare-heated coronal plasma, depends on additional factors; the mapping from proxy to flare occurrence is therefore only probabilistic, with substantial overlap between flaring and flare-quiet regions \citep{lekabarnes2003, barnes2016}. Second, the sunspot number is a count subject to observer-dependent calibration and has undergone major systematic revisions \citep{clette2014, clette2016}. Third, active-region magnetic parameters such as SHARP exist only since the launch of SDO in 2010, covering less than one and a half solar cycles, and describe only regions on the visible disc \citep{bobra2014, pesnell2012sdo}. Fourth, forecasting flares through a proxy is a two-stage process (proxy forecast, then proxy-to-flare conversion) in which the errors of both stages compound. In this work we instead model the flare record measured by the GOES XRS directly \citep{garcia1994, woods2024}. The only sunspot-derived input to our model is the epoch of past solar-cycle minima, used purely as a timing reference for the cycle phase (Sect.~\ref{sec:phase_enc}); the amplitude and shape of the cycle are learned from the flare counts themselves.

\subsection{The flare classification scheme}
\label{sec:classes}

Solar flares are classified by their peak soft X-ray irradiance in the 0.1--0.8~nm (1--8~\AA) band, expressed in $\mathrm{W\,m^{-2}}$ and measured by the XRS channel B on the GOES satellites \citep{garcia1994, machol_guide}:
\begin{itemize}
    \item \textbf{A-class:} near-background emission ($10^{-8}$--$10^{-7}\,\mathrm{W\,m^{-2}}$);
    \item \textbf{B-class:} minor events ($10^{-7}$--$10^{-6}\,\mathrm{W\,m^{-2}}$);
    \item \textbf{C-class:} small events ($10^{-6}$--$10^{-5}\,\mathrm{W\,m^{-2}}$);
    \item \textbf{M-class:} moderate events ($10^{-5}$--$10^{-4}\,\mathrm{W\,m^{-2}}$);
    \item \textbf{X-class:} major events ($\geq 10^{-4}\,\mathrm{W\,m^{-2}}$), with no upper bound.
\end{itemize}
Each successive letter denotes a tenfold increase in peak irradiance, and within a class a numerical suffix gives the multiplicative factor within that decade: an M5.2 flare has a peak irradiance of $5.2\times10^{-5}\,\mathrm{W\,m^{-2}}$ and an X9.0 flare $9.0\times10^{-4}\,\mathrm{W\,m^{-2}}$. The suffix is obtained by truncating, not rounding, the peak irradiance \citep{machol_guide}. The flare catalogue used here defines its class thresholds from B-class upwards \citep{lucas_readme}; A-class emission is therefore not part of our analysis.

\subsection{Class imbalance and observational bias}
\label{sec:imbalance}

\begin{figure}[H]
    \centering
    \includegraphics[width=\linewidth]{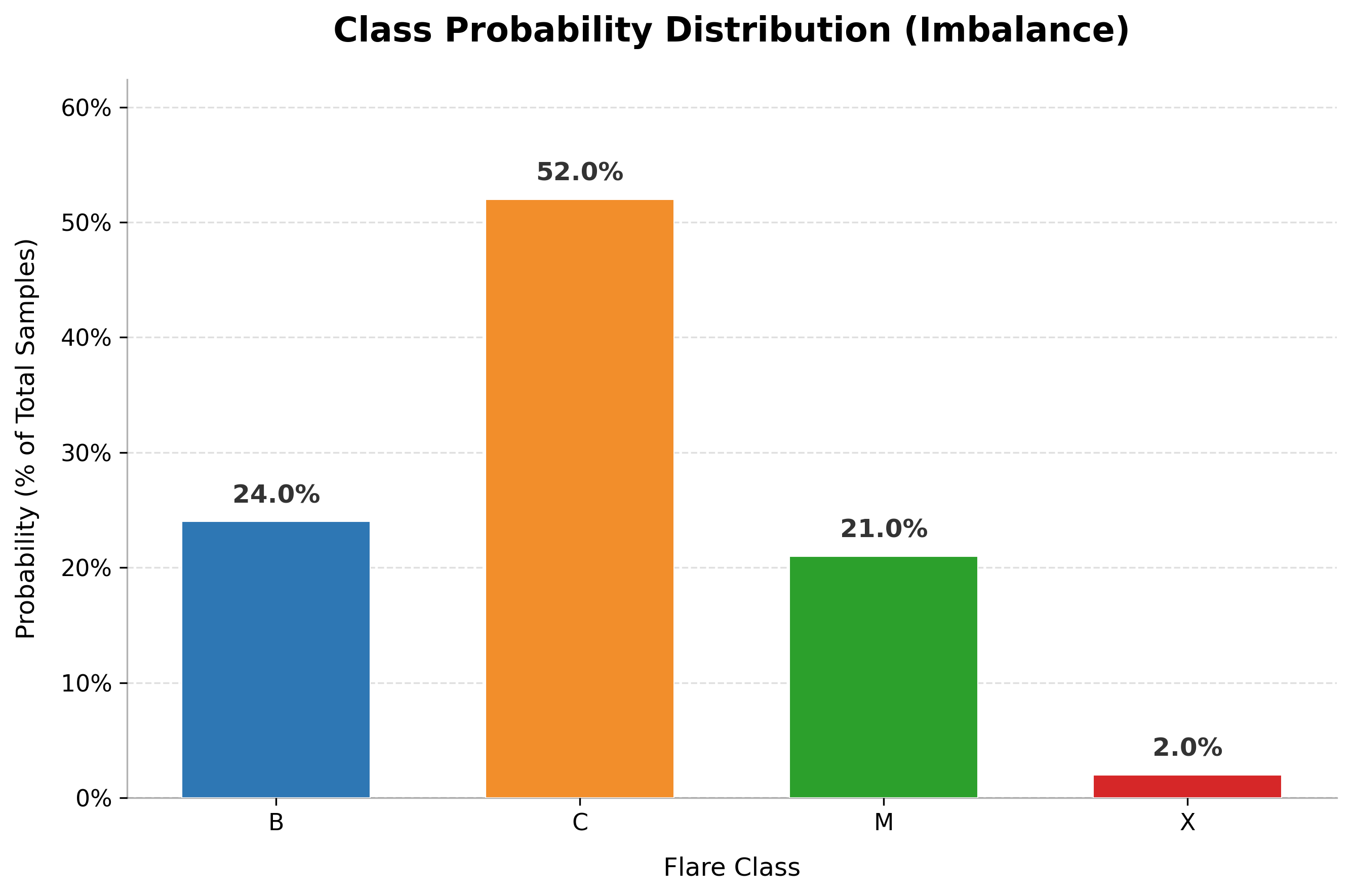}
    \caption{Distribution of solar flares by class in the GOES Flare Report, 1995--2025. The number of events decreases steeply with increasing class, so that M- and X-class events form a small minority of the sample.}
    \label{fig:imbalance}
\end{figure}

The flares in the GOES Flare Report \citep{goes_flrpt} are distributed very unevenly across classes (Fig.~\ref{fig:imbalance}): the number of events falls steeply with increasing class, so that M- and X-class flares together form only a small minority of the catalogue. Such class imbalance is a well-known difficulty for machine-learning classifiers \citep{he2009, angryk2020}.

That this skew is predominantly physical rather than instrumental follows from the flare-size distribution. The number of flares per unit peak flux follows a power law over several orders of magnitude, a property found consistently for soft and hard X-ray flares, over three solar cycles, and with different instruments \citep{hudson1991, crosby1993, aschwanden2012}, and interpreted in the framework of self-organised criticality \citep{aschwanden2012}. Instrumental effects nevertheless contribute at the low-flux end: the detection of weak flares depends on the estimated pre-flare background, which can be biased by noise or by rapidly repeating flares \citep{machol_guide}.

The imbalance creates a formidable modelling challenge for event classification. A classifier that predicts ``C-class'' at every time step would achieve a high nominal accuracy while being operationally worthless for the events that matter. Likewise, a classifier that never predicts an X-class flare would also be highly accurate, yet it is precisely the X-class events that cause the greatest societal impact. Accuracy is therefore an inappropriate figure of merit, and base-rate-independent skill scores such as the true skill statistic are preferred \citep{bloomfield2012}.

\section{This Work:}
\label{sec:thiswork}

We introduce SHAPE, a two-phase machine-learning pipeline for solar flare forecasting. Phase 1 forecasts monthly flare counts, aggregated to annual totals, and thereby captures the $\sim$11-year Schwabe cycle \citep{schwabe1844, hathaway2015}. It combines (i) a physically anchored cycle-shape curve, which fixes the timing and mean profile of each cycle from its phase relative to published solar minima, with (ii) a compact network built from a gated recurrent unit \citep[GRU;][]{cho2014} and a small feedforward branch \citep{goodfellow2016}, which models the residual variability around that curve from the preceding 24 months of residuals and five static inputs: four Fourier terms of the cycle phase and a geomagnetic precursor (Sect.~\ref{sec:arch}). Phase 2 forecasts the strongest flare class expected on the following day (C, M or X) from a compact, shared-trunk LSTM \citep{hochreiter1997} with two output heads, operating on a 12-day rolling window of five engineered features derived from the daily record of C-, M- and X-class flares; its multiclass output can be blended with a deterministic score computed from active-region physical parameters, and a hand-set threshold rule converts the resulting probabilities into an operational class forecast. Phase 1 incorporates Monte Carlo (MC) dropout for principled uncertainty quantification \citep{gal2016}; Phase 2's probability forecasts are instead assessed directly for calibration (Sect.~\ref{sec:phase2_results}).

The paper is structured as follows: Sect.~\ref{sec:data} describes the data; Sect.~\ref{sec:arch} details the neural architectures and reports the Phase 1 results; Sect.~\ref{sec:application} discusses operational applications; Sect.~\ref{sec:future} outlines future directions; and Sect.~\ref{sec:conc} concludes.

\section{Data}
\label{sec:data}

\subsection{The GOES Flare Report}
\label{sec:sources}

The primary data set is the GOES Flare Report, a science-quality composite catalogue of solar flares produced by the National Oceanic and Atmospheric Administration (NOAA) National Centers for Environmental Information (NCEI) from GOES XRS measurements \citep[version 1-0-1;][]{lucas_readme, goes_flrpt}. It covers GOES-8 through GOES-19 from January 1995 onwards and provides one record per flare, with start, peak and end times, the peak XRS-B irradiance and the flare class. Extension of the report to the earlier GOES-1 to GOES-7 satellites, and hence back to 1975, is planned but not yet available \citep{lucas_readme}; our analysis therefore spans 1995--2025, i.e.\ the complete Solar Cycles 23 and 24, the tail of Cycle 22 and the rising and maximum phases of Cycle 25.

The catalogue is derived from the Level-2 (L2) science-quality flare summary, produced by a flare-detection algorithm applied to 1-minute averages of the 0.1--0.8~nm channel \citep{machol_guide}. The underlying XRS cadence is 2--3~s for GOES-8 to GOES-15 and 1~s for the GOES-R series (GOES-16 onwards; \citealt{machol_guide, woods2024}); because the catalogue is built from 1-minute averages, its time resolution is 1~minute throughout the record and is unaffected by this change of cadence \citep{lucas_readme}. At least two GOES satellites operate at any time. For each flare the record of the designated primary satellite is used when available, and otherwise that of the secondary; events seen by two satellites within 4~minutes and within $\pm15\%$ in peak irradiance are treated as the same flare \citep{lucas_readme}. This redundancy, together with the fact that gaps in one satellite's record are filled from the other, makes the catalogue considerably less sensitive to single-instrument outages than earlier single-satellite lists.

In contrast to the operational archive, the science-quality products incorporate retrospective fixes for known instrument issues and outages, recover data that were missing in real time, and use the most recent calibration \citep{machol_guide}. In particular, the scaling factors that the NOAA Space Weather Prediction Center (SWPC) applied to GOES-8 to GOES-15 irradiances (0.7 for XRS-B) have been removed, so that irradiances are in true physical units and flare classes are consistent across all satellites; an X2.5 flare reported operationally by GOES-15 corresponds to an X3.6 flare on the GOES-R scale \citep{machol_guide, lucas_readme}. Saturation of the flare peak is flagged and has occurred only four times in the record (2001 April 2, 2003 October 28, 2003 November 4 and 2005 September 7; \citealt{lucas_readme}). Because these issues are treated at source, we apply no additional event-level correction, interpolation or manual insertion of events. Residual caveats remain and users are responsible for inspecting the data \citep{lucas_readme}; these are discussed in Sect.~\ref{sec:limits}.

\subsection{Auxiliary data}
\label{sec:aux}

Two auxiliary inputs are used (Table~\ref{tab:data}). The first is the dates of the solar-cycle minima, taken from the World Data Center SILSO \citep{silso} and the NOAA/NASA Prediction Panel for the current minimum \citep{swpc_panel2019}, which we use as the anchor of the cycle phase (Table~\ref{tab:minima}). Cycle lengths in this interval range from 9.6 to 12.6~yr with a mean of 10.9~yr, consistent with the general range of cycle lengths reported in the literature \citep{hathaway2015}. The second is the geomagnetic aa index, derived from K-indices at two nearly antipodal mid-latitude observatories \citep{mayaud1972} and distributed as 3-hourly values by the International Service of Geomagnetic Indices \citep[ISGI;][]{isgi}. We use the definitive series for 1995--2024 and the provisional series for 2025, averaged to monthly values. The role of the aa index as a solar-cycle precursor is described in Sect.~\ref{sec:precursor}.

\begin{table}[H]
    \caption{Data used in Phase 1.}
    \label{tab:data}
    \centering
    \small
    \begin{tabular}{llll}
        \hline\hline
        Data set & Content & Span & Role \\
        \hline
        \parbox[t]{2.0cm}{\raggedright GOES Flare Report v1-0-1} & \parbox[t]{2.5cm}{\raggedright Flare events (peak time, class, irradiance)} & \parbox[t]{1.6cm}{\raggedright 1995--2025} & \parbox[t]{1.2cm}{\raggedright Target} \\[4pt]
        \parbox[t]{2.0cm}{\raggedright aa index (ISGI)} & \parbox[t]{2.5cm}{\raggedright 3-hourly geomagnetic index; monthly means} & \parbox[t]{1.6cm}{\raggedright 1995--2024 (definitive), 2025 (prov.)} & \parbox[t]{1.2cm}{\raggedright Static precursor} \\[4pt]
        \parbox[t]{2.0cm}{\raggedright Cycle minima (SILSO)} & \parbox[t]{2.5cm}{\raggedright Dates of solar-cycle minima} & \parbox[t]{1.6cm}{\raggedright Cycles 21--25} & \parbox[t]{1.2cm}{\raggedright Phase anchor} \\
        \hline
    \end{tabular}
\end{table}

\begin{table}[H]
    \caption{Solar-cycle minima used as phase anchors. Entries marked with an asterisk are extrapolated using the mean cycle length of 10.87~yr and are not observed.}
    \label{tab:minima}
    \centering
    \small
    \begin{tabular}{lccc}
        \hline\hline
        Cycle & Start minimum & Decimal year & Length (yr) \\
        \hline
        21 & 1976 Jun & 1976.50 & 10.25 \\
        22 & 1986 Sep & 1986.75 & 9.60 \\
        23 & 1996 May & 1996.35 & 12.61 \\
        24 & 2008 Dec & 2008.96 & 11.00 \\
        25 & 2019 Dec & 2019.96 & (ongoing) \\
        \hline
    \end{tabular}
\end{table}

\subsection{Aggregation and annual series}
\label{sec:aggregation}

\begin{figure*}[t]
    \centering
    \includegraphics[width=0.80\textwidth]{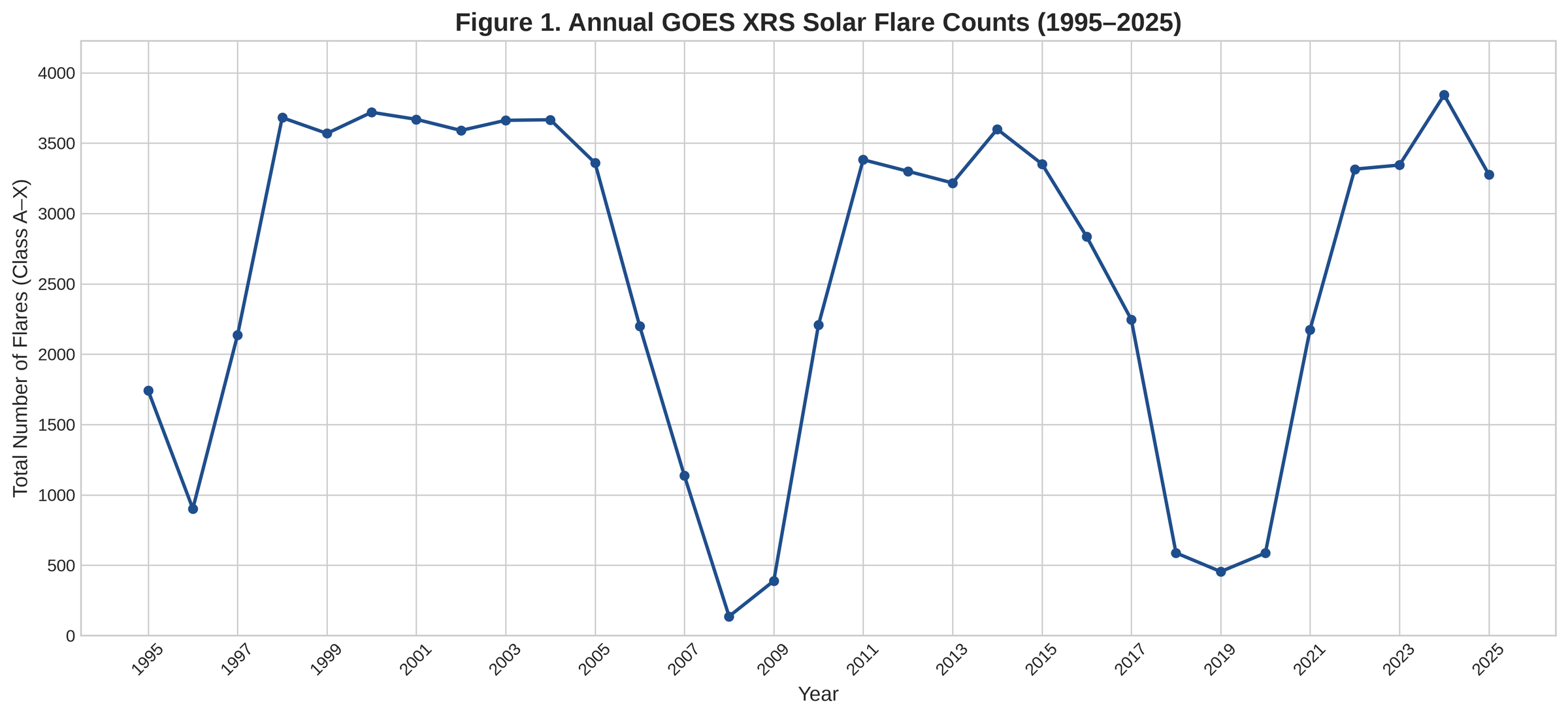}
    \caption{Annual number of flares in the GOES Flare Report (all classes listed in the catalogue), 1995--2025.}
    \label{fig:input}
\end{figure*}

Flares are counted by peak time in calendar months to give the monthly series $N(t)$. Months without a recorded flare, which occur near deep minima, are retained as zero rather than dropped. Only complete calendar years are used: the record extends into 2026, but 2026 is excluded as incomplete, leaving 372 monthly and 31 annual values for 1995--2025. Annual totals (Fig.~\ref{fig:input}) are obtained by summing the twelve monthly values of each calendar year. The series shows the solar cycle clearly, with minima in 1996, 2008--2009 and 2018--2020, when fewer than $\sim$600 flares were recorded per year (fewer than 200 in 2008), and with maxima of $\sim$3.3--3.9$\times10^{3}$ flares per year around 1998--2004, 2011--2015 and 2022--2025.

\section{Neural architecture}
\label{sec:arch}

\subsection{Phase 1: Annual flare count regression}
\label{sec:phase1}

\subsubsection{Design rationale and overview}

We have only 31 annual flare totals (one for each year from 1995 to 2025).
This is far too little data for a recurrent neural network, which would
simply memorise it (overfit). The catalogue, however, records the time of
every flare, so we can count flares month by month instead. This gives
372 monthly values (31 years $\times$ 12 months), which is enough to train
a small network, provided that we take two precautions. First, the network
must be \emph{compact}, meaning it has few adjustable parameters (weights)
and therefore little room to memorise the data. Second, it must be
\emph{strongly regularised}, meaning that penalties and random dropout of
network units are applied during training to stop it from fitting noise
\citep{goodfellow2016}. Both are described in Sect.~\ref{sec:arch}.

\begin{figure*}[t]
    \centering
    \includegraphics[width=0.80\textwidth]{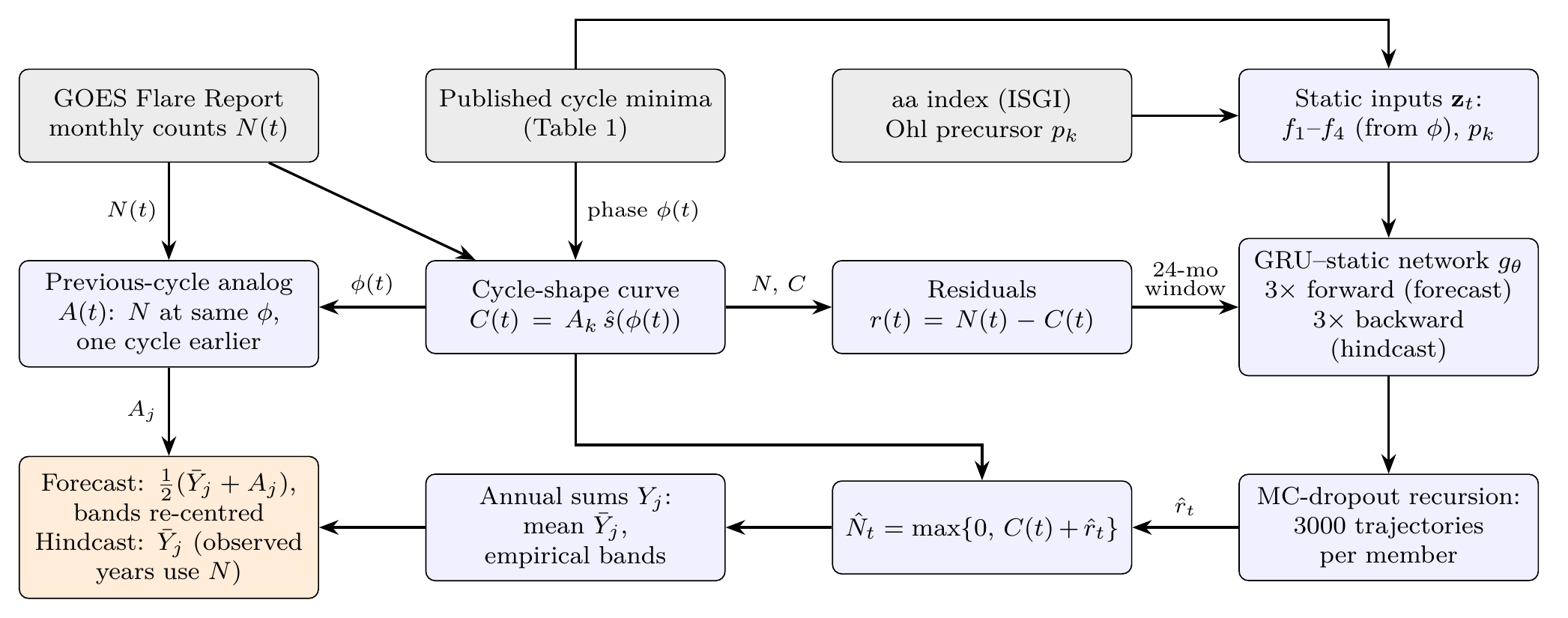}
    \caption{Overview of the Phase~1 pipeline. The cycle-shape curve $C(t)$ is computed solely from the cycle phase. Residuals are formed from both $N(t)$ and $C(t)$. Only the residual is forecast recursively; $C(t)$ is never fed back.}
    \label{fig:pipeline_phase_1}
\end{figure*}

We split the monthly flare count $N(t)$ into two parts:
\begin{equation}
    N(t) = C(t) + r(t).
\end{equation}
Here $C(t)$ is the smooth, expected number of flares in a month, given
where that month falls in the solar cycle. It is a fixed curve with no
randomness (deterministic). The residual $r(t)$ is the difference between
the actual count and this curve, i.e.\ how far a given month lies above or
below it. The neural network forecasts only $r(t)$
(Fig.~\ref{fig:pipeline_phase_1}). For example, if the curve gives 300
flares for a month and we count 340, the residual is $+40$.

Also see Fig.~\ref{fig:pipeline_phase_1}. This separation is deliberate. Forecasting many months ahead by feeding a network's own predictions back into itself accumulates error and typically produces forecasts that lag turning points \citep{taieb2012}. With the decomposition above, the timing and amplitude of the cycle come from $C(t)$, which is never fed back, and only the residual is forecast recursively, so that its drift stays centred on the curve rather than on zero. The model predicts flare counts month by month. We add up the twelve monthly values in each year to get annual totals, and we compare these with the observed annual counts.

\subsubsection{Cycle-phase encoding}
\label{sec:phase_enc}

The dominant variability of solar activity is governed by the Schwabe cycle, with a mean period of about 11 years \citep{schwabe1844, hathaway2015}. Individual cycles, however, vary considerably in length: 9.6 to 12.6~yr in our sample (Table~\ref{tab:minima}). A phase defined with a fixed period from a fixed epoch would therefore drift by about a year within one or two cycles, and any forecast built on it would be progressively mistimed. We instead anchor the phase to the published minima $\tau_k$ and define, for $\tau_k \le t < \tau_{k+1}$,
\begin{equation}
    \phi(t) = \frac{t-\tau_k}{\tau_{k+1}-\tau_k}\in[0,1),
\end{equation}
so that the phase is reset at every observed minimum. Beyond the last observed minimum (2019.96) the anchors are extrapolated with the mean cycle length of 10.87~yr (Table~\ref{tab:minima}); the timing of Cycles 26 and 27 is therefore an assumption and not a prediction (Sect.~\ref{sec:limits}). The phase enters the network through four Fourier features,
\begin{align}
    f_1 &= \sin(2\pi\phi), & f_2 &= \cos(2\pi\phi), \nonumber\\
    f_3 &= \sin(4\pi\phi), & f_4 &= \cos(4\pi\phi).
\end{align}
Higher harmonics of sin and cosine are included to capture the systematic difference in the rising and decaying rates of the solar cycle, and an occasional double peaked maxima eg.
\citep{waldmeier1935, hathaway1994} \st{and often display a double peak separated by the Gnevyshev gap} \citep{gnevyshev1977}, Moreover the explicit Fourier features are also known to help networks fit high-frequency functions of low-dimensional inputs \citep{tancik2020}.

\subsubsection{Cycle-shape curve}

The mean shape of the cycle is learned from the completed cycles in the
training data. A cycle is complete when both of its bounding minima lie
inside the training interval. With training data from 1995 to 2020, these are Cycles 23 and 24. (Cycle 22 begins before the record, and Cycle 25 is not yet finished.) For each completed cycle $k$ we find its peak $P_k$ as follows.
We replace each monthly count $N(t)$ by the average of the 13 months centred on it (6 months before, the month itself, and 6 months
after). This 13-month running mean smooths out month-to-month noise, in the same way that the sunspot number is smoothed \citep{clette2014}. The peak $P_k$ is the highest value of this smoothed series within the cycle. For example, suppose the 13 monthly counts around one month in the maximum of a cycle average to 280 flares. If no other month in that cycle has a higher 13-month mean, then $P_k = 280$ flares per month.\citep{clette2014}. The counts of each cycle are normalised by $P_k$ and distributed in in $20$, $\phi$-bins, giving a template $\hat s(\phi)$ that is interpolated periodically. The curve is then
\begin{equation}
    C(t) = A_{k(t)}\,\hat s\bigl(\phi(t)\bigr),
\end{equation}
where $A_k = P_k$ for a completed cycle and, for any cycle without a completed peak (Cycle 22 in the hindcast and Cycles 25 onwards in the forecast), $A_k$ is the peak of the most recent completed cycle (Cycle 24), i.e.\ a persistence assumption. This is a deliberately simple choice: with amplitude information from only two cycles, no regression of amplitude on a precursor can be calibrated, and persistence is the appropriate reference (Sect.~\ref{sec:limits}). The residual series is $r(t)=N(t)-C(t)$.

\subsubsection{Geomagnetic precursor}
\label{sec:precursor}

The minimum of the smoothed aa index near a solar minimum is correlated with the amplitude of the following cycle \citep{ohl1966, feynman1982, thompson1993}, and geomagnetic precursors are among the best-established tools of cycle prediction \citep{petrovay2020}. For every cycle boundary we compute the minimum of the 13-month smoothed monthly aa index within $\pm1.5$~yr of the published minimum, using only data from before the training cutoff, and use it as a per-cycle static input $p_k$. Cycles for which the aa record does not reach the corresponding minimum receive the median of the available values. The aa record begins in 1995, so only three minima (1996, 2008 and 2019) are covered. This is too few to regress cycle amplitude on the precursor; $p_k$ therefore enters the residual network only as a weakly informative static input, and its contribution should not be given a physical interpretation. A longer aa record, available from 1868 \citep{mayaud1972, isgi}, would allow $A_k$ to be regressed on $p_k$ following \citet{petrovay2020}.

\subsubsection{Network architecture}

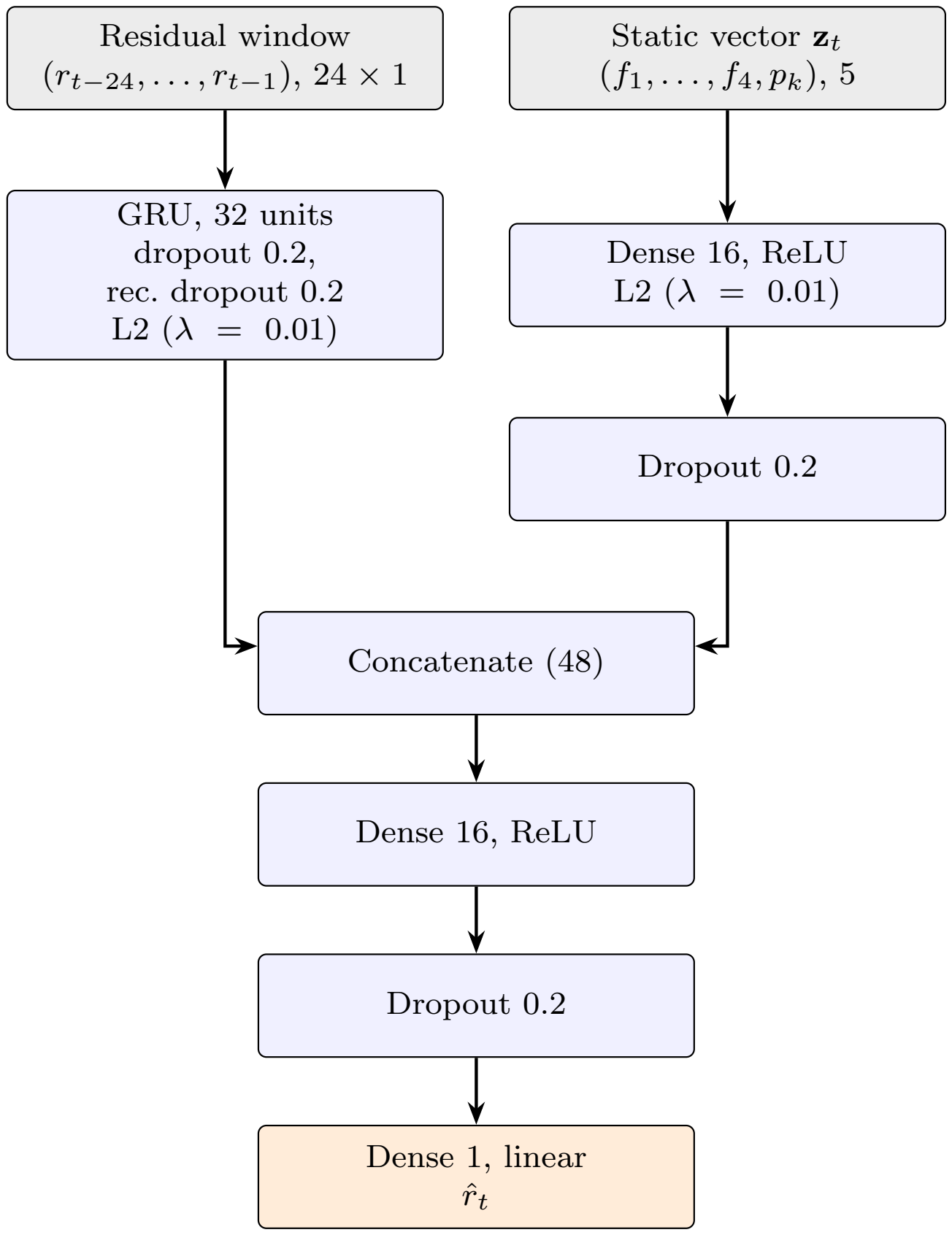
\begin{figure}[H]
\centering
\resizebox{\columnwidth}{!}{%
\begin{tikzpicture}[
  box/.style={draw, rounded corners=2pt, align=center, font=\footnotesize, text width=3.6cm, minimum height=0.9cm, fill=blue!6},
  inp/.style={box, fill=gray!15},
  outp/.style={box, fill=orange!15},
  arr/.style={-{Stealth[length=2mm]}, thick}]
\node[inp] (in1) at (0,0)   {Residual window\\ $(r_{t-24},\ldots,r_{t-1})$, $24\times1$};
\node[inp] (in2) at (4.4,0) {Static vector $\mathbf z_t$\\ $(f_1,\ldots,f_4,p_k)$, 5};
\node[box] (gru) at (0,-1.9) {GRU, 32 units\\ dropout 0.2, rec.\ dropout 0.2\\ L2 ($\lambda=0.01$)};
\node[box] (d1)  at (4.4,-1.9) {Dense 16, ReLU\\ L2 ($\lambda=0.01$)};
\node[box] (do1) at (4.4,-3.6) {Dropout 0.2};
\node[box] (cat) at (2.2,-5.3) {Concatenate (48)};
\node[box] (d2)  at (2.2,-6.8) {Dense 16, ReLU};
\node[box] (do2) at (2.2,-8.3) {Dropout 0.2};
\node[outp] (out) at (2.2,-9.8) {Dense 1, linear\\ $\hat r_t$};
\draw[arr] (in1) -- (gru);
\draw[arr] (in2) -- (d1);
\draw[arr] (d1) -- (do1);
\draw[arr] (gru.south) -- ++(0,-0.6) |- ([yshift=0.15cm]cat.west);
\draw[arr] (do1.south) -- ++(0,-0.6) |- ([yshift=0.15cm]cat.east);
\draw[arr] (cat) -- (d2);
\draw[arr] (d2) -- (do2);
\draw[arr] (do2) -- (out);
\end{tikzpicture}}
\caption{Architecture of the Phase 1 residual network.}
\label{fig:arch_phase1}
\end{figure}

The residual $r(t)$ is forecast one month ahead by a compact two-branch network (Fig.~\ref{fig:arch_phase1}):
\begin{equation}
    \hat r_t = g_\theta\bigl(r_{t-24},\ldots,r_{t-1};\ \mathbf z_t\bigr),\qquad \mathbf z_t=(f_1,f_2,f_3,f_4,p_{k(t)}).
\end{equation}

A recurrent branch reads the preceding $W=24$ monthly residuals; a feedforward branch processes the static vector $\mathbf z_t$; and the two are merged. The layers are:
\begin{enumerate}
    \item \textbf{Recurrent branch}: GRU \citep{cho2014} with 32 units, input dropout 0.2 and recurrent dropout 0.2, with L2 regularisation ($\lambda=0.01$) on the input weights \citep{krogh1992}.
    \item \textbf{Static branch}: dense layer with 16 units, ReLU activation \citep{nair2010} and L2 regularisation ($\lambda=0.01$), followed by dropout with rate 0.2 \citep{srivastava2014}.
    \item \textbf{Merge}: concatenation of the two branches (48 features), followed by a dense layer with 16 units and ReLU activation and dropout with rate 0.2.
    \item \textbf{Output}: a single linear neuron returning $\hat r_t$.
\end{enumerate}
The network has about $4.3\times10^{3}$ trainable parameters. The architecture reflects the small sample of 288 training windows. A GRU has fewer parameters than an LSTM with comparable performance on many sequence tasks \citep{chung2014}, which is important at this sample size; the two-branch design separates the deterministic phase information from the stochastic recent history, so that each is processed by a layer type suited to it; and the same L2 penalty is applied to both branches so that neither is favoured. Dropout in the recurrent branch is applied to the recurrent connections as well, since this is the theoretically grounded form for MC dropout in recurrent networks \citep{gal2016rnn}.

\subsubsection{Training protocol}

The model is trained on months up to and including December 2020 (312 months, giving 288 windows), so that the years 2021--2025 are never seen. The residual series and the static features are scaled to $[0,1]$ with min-max scalers fitted on the training data \citep{pedregosa2011}; the same scaler is used for the input residuals and the target so that recursive predictions are fed back on the same scale. We use the Adam optimiser \citep{kingma2015} with a mean-squared-error loss, a batch size of 16, and TensorFlow \citep{abadi2016}.

The training length is chosen by early stopping \citep{prechelt1998} on the rising phases of the cycle. The rising phase, taken as the first four years after each minimum, is the part of the cycle in which the forecast must change most quickly, whereas a conventional validation split of the final rows would hold out only the decline and minimum. Each fold holds out the target months of one rising phase (those of Cycles 23 and 24), the network is trained on the remaining rows with a patience of 25 epochs and a maximum of 300, and the best epochs of the folds are averaged. The final networks are then trained on all rows for that fixed number of epochs. Three networks with different random seeds are trained, forming an ensemble \citep{lakshminarayanan2017}. A second ensemble is trained on the time-reversed series and is used only for the hindcast.

\subsubsection{Uncertainty quantification and forecast generation}

Predictive uncertainty is estimated by MC dropout \citep{gal2016}: dropout remains active at inference, so that each forward pass samples a different sub-network. Within each pass the recurrent dropout masks are shared across the 24 steps of the input window \citep{gal2016rnn}, and new masks are drawn at every recursive step. Starting from the last 24 observed residuals, each trajectory $m$ is advanced month by month for 252 months (2021 January to 2041 December), each predicted residual being appended to the input window, and the count trajectory is
\begin{equation}
    \hat N^{(m)}_t = \max\bigl\{0,\; C(t)+\hat r^{(m)}_t\bigr\},
\end{equation}
which enforces non-negative counts. We generate 3000 trajectories for each of the three ensemble members and pool the resulting 9000 trajectories. Annual totals $Y_j^{(m)}$ are obtained by summing the twelve monthly values of each year. The forecast is the mean over trajectories and the uncertainty bands are the empirical percentiles: for a confidence level $c$ the band is bounded by the $100(1-c)/2$ and $100(1+c)/2$ percentiles of $\{Y_j^{(m)}\}$. We show $c=0.6827$ and $c=0.9973$. Percentiles are used instead of a Gaussian mean $\pm z\sigma$ because counts are non-negative and generally skewed. The hindcast for 1985-1994 is generated in the same way by the backward ensemble, seeded with the earliest 24 residuals in reverse order; for 1995 and 1996, which are observed, the observed values are shown.

\subsubsection{Baselines and evaluation protocol}

The hybrid model must outperform two parameter-free baselines that use only training data. The first is the shape curve $C(t)$ on its own, i.e.\ without the residual network. The second is the previous-cycle analogue \citep{lorenz1969}, which predicts the count at the same phase of the previous cycle. The forecast shown in Fig.~\ref{fig:output_phase1} is the mean of the hybrid ensemble forecast and the previous-cycle analogue. 
Skill is measured on the held-out years 2021-2025 by the mean absolute error (MAE), the root-mean-square error (RMSE) and the fraction of years whose observed value lies in the 99.73\% band. In addition, we perform a rolling-origin evaluation \citep{tashman2000} in which the entire pipeline (cycle peaks, template, aa precursor, epochs) is rebuilt using only data before a cutoff at a cycle boundary: a cutoff at the Cycle 23/24 minimum (end of 2008) forecasting 2009-2014, and a cutoff at the Cycle 24/25 minimum (end of 2019) forecasting 2020-2025. The back-tests use a single network and 1500 trajectories. The 2008 back-test has a single completed cycle available for the template.

\subsubsection{Results}
\label{sec:results}

\begin{figure*}[t]
    \centering
    \includegraphics[width=0.95\textwidth]{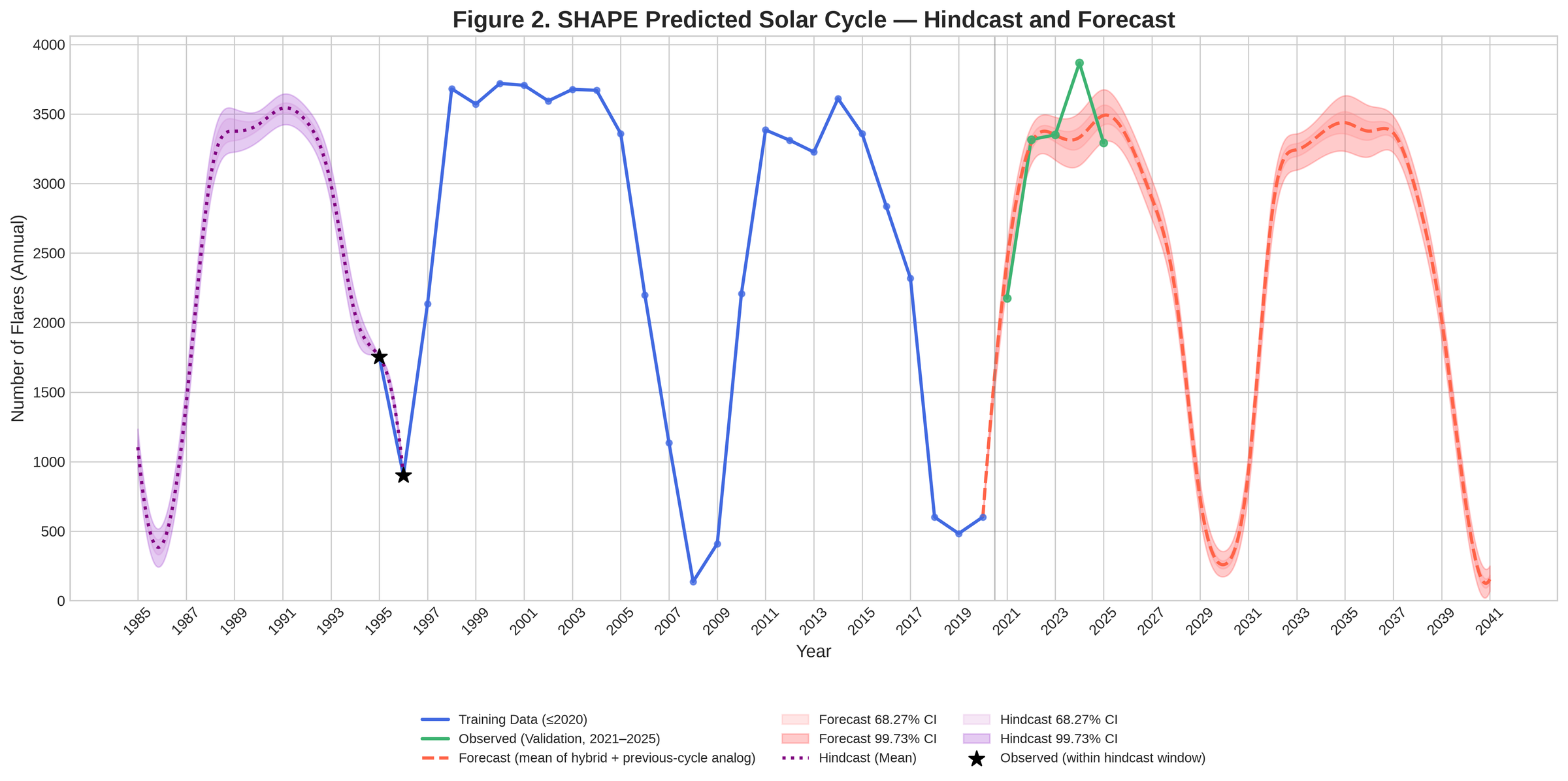}
    \caption{Phase 1 output: hindcast and forecast of the annual flare count. Blue: training data ($\leq2020$); green: observed values for 2021-2025, withheld from training; dashed red: forecast (mean of the hybrid ensemble and the previous-cycle analogue) with 68.27\% and 99.73\% empirical bands; dotted purple: hindcast with the same bands, where the stars mark the observed years 1995-1996. Curves are cubic-spline interpolations of annual values.}
    \label{fig:output_phase1}
\end{figure*}

Trained on data through December 2020, the model was evaluated on the five subsequent years, which it never saw (Fig.~\ref{fig:output_phase1}). The forecast reproduces the steep rise out of the 2020 minimum and the subsequent high-activity plateau of $\sim$3.3-3.5$\times10^{3}$ flares per year. The observed counts of 2022 and 2023 lie very close to the forecast mean, whereas 2024, the year of the observed maximum, was underestimated and lies above the upper 99.73\% band; the counts of 2021 and 2025 are slightly overestimated. Over 2021-2025 the MAE is 204 flares yr$^{-1}$ and the RMSE 280 flares yr$^{-1}$, and 4 of 5 of the five observed years lie within the 99.73\% band. This limitation is directly relevant to the underestimation of the 2024 maximum noted above. The Phase~1 model is trained only on the historical flare-count time series; it takes no contemporaneous measure of sunspot or active-region number as an input, and so cannot anticipate a departure from the learned cycle-shape template that is driven by an unusually active photosphere. The year 2024 was independently confirmed as the maximum of Cycle~25, with the smoothed sunspot number peaking at $\sim$161 in October 2024 \citep{sidc2025} and the (non-smoothed) monthly sunspot number reaching its highest value in over two decades \citep{swpc_update2023}. This was accompanied by an unusually large number of large, complex active regions crossing the disk through 2024 \citep{sidc2025}, in contrast to the comparatively quiet Cycle 24 maximum. Because flare occurrence is well established to track the sunspot and active-region number closely, with reported correlation coefficients of $\sim$0.87-0.99 across multiple cycles \citep{temmer2001, lee2012, joshi2015, aschwanden2021}, and because active regions with more complex ($\delta$-type) magnetic configurations are disproportionately flare-productive \citep{sammis2000, kumar2025}, the surge in active-region number and complexity in 2024 plausibly explains why the observed flare count exceeded both the cycle-shape template and the model's upper 99.73\% band for that year. A model incorporating a contemporaneous active-region or sunspot-number covariate, rather than flare counts alone, may therefore better anticipate such short-term departures from the smooth cycle envelope; we return to this in Sect.~\ref{sec:future}. 

Table~\ref{tab:skill} compares the hybrid with the two baselines and reports the rolling-origin back-tests.

\begin{table}[H]
    \caption{Mean absolute error (flares yr$^{-1}$) of the hybrid model and the two baselines on the held-out years and in the rolling-origin back-tests.}
    \label{tab:skill}
    \centering
    \small
    \begin{tabular}{lccc}
        \hline\hline
        Model & 2021-2025 &  2008 &  2019 \\
        \hline
        Hybrid ensemble & 235 & 453 & 238 \\
        Shape curve only & 259 & 503 & 265  \\
        Previous-cycle analogue & 197 & 501 & 191 \\
        \hline
    \end{tabular}
\end{table}

Beyond the validation period, the mean forecast declines towards a minimum near 2030, rises again in 2031-2032, and forms a plateau of $\sim$3.3-3.4$\times10^{3}$ flares per year over 2032-2037, before declining towards the next minimum at the end of the horizon. The hindcast recovers the minimum of 1986 and a maximum in 1989-1991, which agrees with the timing of Cycle 22 \citep{hathaway2015}; it cannot, however, be validated against flare data with the present catalogue, because the flare report currently begins in 1995. An independent test will become possible once the report is extended to the earlier GOES satellites \citep{lucas_readme}.

\subsubsection{Limitations}
\label{sec:limits}

Several limitations should be borne in mind when interpreting these results. (i)~The amplitude of every future cycle is taken from Cycle 24 (persistence), and the template rests on only two completed cycles; the model is therefore not an independent forecast of the amplitude of Cycle 25 or 26. (ii)~The timing of Cycles 26 and 27 is fixed by the mean cycle length, whereas observed lengths span 9.6-12.6~yr (Table~\ref{tab:minima}), so a timing error of one to two years is plausible. (iii)~The uncertainty bands represent the variability of the network (dropout and ensemble initialisation) conditional on the curve $C(t)$; they do not include uncertainty in amplitude, template or phase, and are hence too narrow, as the 2024 maximum illustrates. (iv)~The aa precursor is covered by only three cycle minima (Sect.~\ref{sec:precursor}).

(v)~The annual totals of all flares appear to saturate: the maxima of Cycles 23, 24 and 25 all lie at $\sim$3.6-3.9$\times10^{3}$ flares per year (Fig.~\ref{fig:input}), differing by less than $\sim$10\%, although their smoothed sunspot-number maxima differ by more than 50\% (about 180, 116 and 161, respectively; \citealt{stce_sc25, swpc_update2023, sidc2025}). Whether this reflects physics or the behaviour of flare detection at high activity, where overlapping flares and elevated backgrounds complicate event identification \citep{machol_guide}, cannot be decided from total counts alone; it limits the sensitivity of the total count to cycle strength and motivates class-resolved analysis. (vi)~The hindcast before 1995 is unvalidated (Sect.~\ref{sec:results}).

\subsection{Phase 2: Daily flare-class forecasting}
\label{sec:phase2}
\subsubsection{Data preprocessing and feature engineering}
\label{sec:phase2_data}

Phase 2 operates on the same GOES Flare Report (Sect.~\ref{sec:sources}) as Phase 1, but at the level of individual events rather than monthly totals, and restricted to a subset of classes. Every event-level record is first reduced to its single-letter class (C, M or X); A- and B-class flares are dropped entirely; and, for calendar days on which more than one C-, M- or X-class flare occurred, only the strongest is retained. The result is a daily series of at most one labelled event per day, but with an important consequence: days on which the strongest flare was A- or B-class, or on which no flare was recorded at all, do not appear in the series as zero-activity days but are simply absent. The 12-day windows used below (and the lagged/rolling features that feed them) are therefore built over the sequence of C/M/X event-days, not over a uniform calendar-day grid; a ``12-day window'' can span anywhere from 12 calendar days, during an active period, to several weeks during a quiet one. We return to this in Sect.~\ref{sec:phase2_limits}.

Each retained event is mapped to an ordinal intensity,
\begin{equation}
    I(c) = \begin{cases} 10, & c=\text{C}\\ 100, & c=\text{M}\\ 1000, & c=\text{X}, \end{cases}
\end{equation}
and four further features are derived from this intensity series: the three most recent values ($\mathrm{Lag}_1,\mathrm{Lag}_2,\mathrm{Lag}_3$) and a 7-event rolling maximum. The five-dimensional feature vector at event-day $t$ is thus $\mathbf x_t = (I_t, \mathrm{Lag}_{1,t}, \mathrm{Lag}_{2,t}, \mathrm{Lag}_{3,t}, \mathrm{RollMax}_{7,t})$. Features are standardised (zero mean, unit variance) with a scaler fitted only on the training split (Sect.~\ref{sec:phase2_train}) \citep{pedregosa2011}, and consecutive feature vectors are grouped into overlapping windows of length 12 to form the model input.

\subsubsection{Network architecture}
\label{sec:phase2_arch}

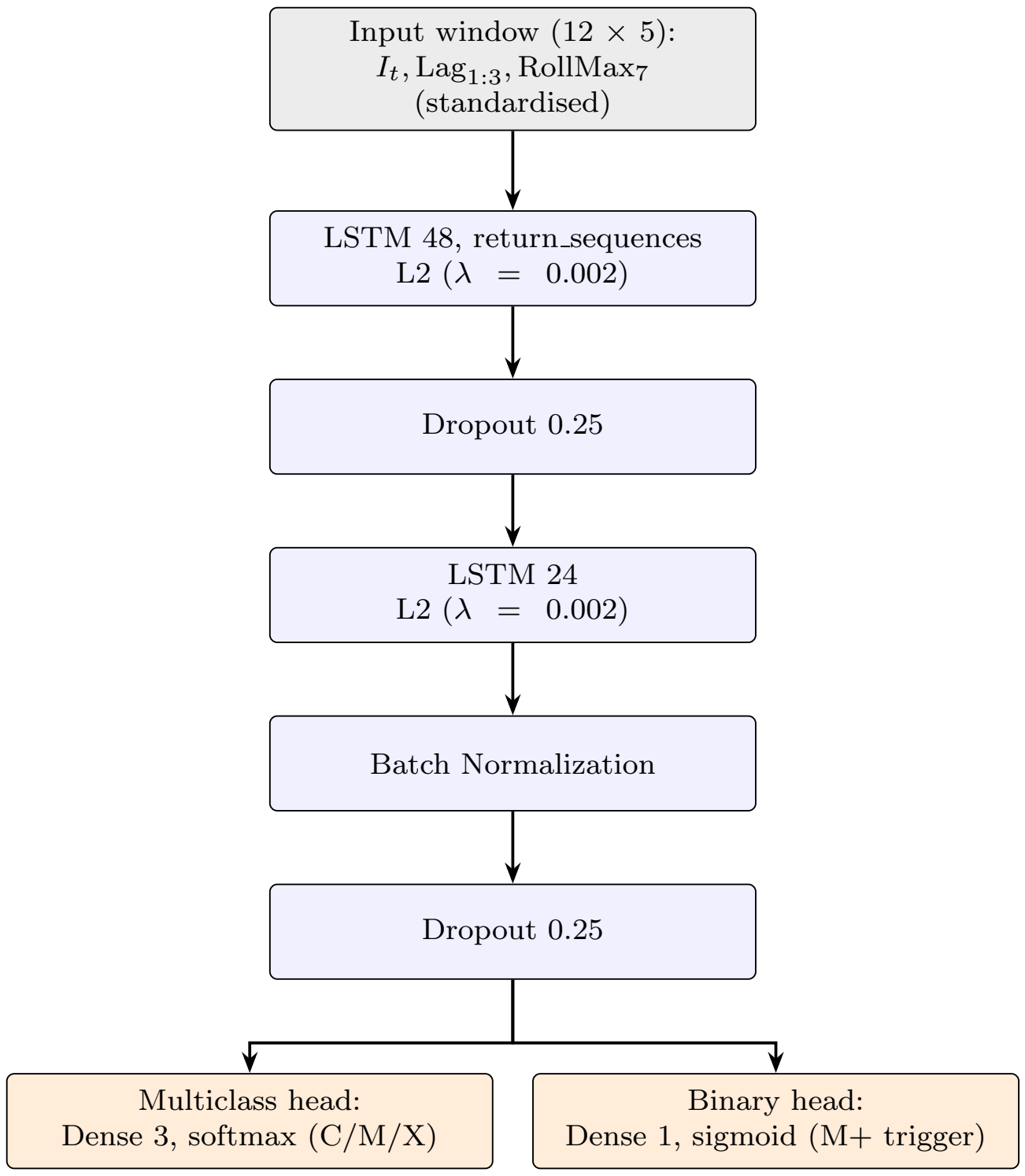
\begin{figure}[H]
\centering
\resizebox{\columnwidth}{!}{%
\begin{tikzpicture}[
  box/.style={draw, rounded corners=2pt, align=center, font=\footnotesize, text width=4.4cm, minimum height=0.9cm, fill=blue!6},
  inp/.style={box, fill=gray!15},
  outp/.style={box, fill=orange!15},
  arr/.style={-{Stealth[length=2mm]}, thick}]
\node[inp] (in1) at (0,0)    {Input window $(12\times5)$:\\ $I_t,\mathrm{Lag}_{1:3},\mathrm{RollMax}_7$ (standardised)};
\node[box] (l1)  at (0,-1.8) {LSTM 48, return\_sequences\\ L2 ($\lambda=0.002$)};
\node[box] (d1)  at (0,-3.4) {Dropout 0.25};
\node[box] (l2)  at (0,-5.0) {LSTM 24\\ L2 ($\lambda=0.002$)};
\node[box] (bn)  at (0,-6.6) {Batch Normalization};
\node[box] (d2)  at (0,-8.2) {Dropout 0.25};
\node[outp] (h1) at (-2.5,-10.0) {Multiclass head:\\ Dense 3, softmax (C/M/X)};
\node[outp] (h2) at (2.5,-10.0)  {Binary head:\\ Dense 1, sigmoid (M+ trigger)};
\draw[arr] (in1) -- (l1);
\draw[arr] (l1) -- (d1);
\draw[arr] (d1) -- (l2);
\draw[arr] (l2) -- (bn);
\draw[arr] (bn) -- (d2);
\draw[arr] (d2.south) -- ++(0,-0.6) -| (h1.north);
\draw[arr] (d2.south) -- ++(0,-0.6) -| (h2.north);
\end{tikzpicture}}
\caption{Architecture of the Phase 2 shared-trunk, dual-head LSTM classifier.}
\label{fig:arch_phase2}
\end{figure}

Phase 2 predicts two related targets from a single shared representation: the flare class itself (C, M or X) and a coarser binary indicator of whether the day is M-or-above (``M+''). Rather than training four separate class-specific classifiers, both targets are produced by two output heads attached to one recurrent trunk, so that the network cannot learn contradictory representations for the two tasks and the harder, rarer M+ decision benefits from the same features as the easier three-way classification \citep[cf.][for related multi-task LSTM designs]{zhang2024}. The trunk and heads are (Fig.~\ref{fig:arch_phase2}):
\begin{enumerate}
    \item \textbf{Input}: a $(12\times5)$ window of the standardised features above.
    \item \textbf{LSTM}: 48 units, \texttt{return\_sequences=True}, L2 regularisation ($\lambda=0.002$) on the recurrent kernel \citep{hochreiter1997, krogh1992}.
    \item \textbf{Dropout}: rate 0.25.
    \item \textbf{LSTM}: 24 units, L2 regularisation ($\lambda=0.002$).
    \item \textbf{Batch normalisation}, followed by \textbf{dropout} at rate 0.25.
    \item \textbf{Multiclass head}: dense layer, 3 units, softmax (C/M/X class probabilities).
    \item \textbf{Binary head}: dense layer, 1 unit, sigmoid (probability the day is M+).
\end{enumerate}
The two heads share every layer up to the final dense layer. The model is trained with a combined loss, sparse categorical cross-entropy for the multiclass head and binary cross-entropy for the binary head, weighted 1.0 and 2.0 respectively, so that errors on the rarer M+ indicator are penalised twice as heavily as multiclass errors during training; this asymmetric weighting is a training-time mechanism only; it does not, by itself, calibrate the inference-time decision rule of Sect.~\ref{sec:phase2_decision}.

\subsubsection{Physics-score blending}
\label{sec:phase2_phys}

A separate, deterministic scoring function converts five active-region parameters (magnetic shear angle, Hale/Mount Wilson magnetic complexity class, vertical field $B_z$, twist number, and free magnetic energy) into a single additive score $S\in[0,5.2]$ by thresholding each parameter against two empirical cut-offs and summing the corresponding point values; $S$ is then mapped to a soft class distribution $p_{\mathrm{phys}}$ by evaluating a Gaussian-shaped kernel centred on three fixed points on the score axis, one per class ($\mu_{\mathrm C}=0.8$, $\mu_{\mathrm M}=2.8$, $\mu_{\mathrm X}=4.5$) with temperature $T=1.2$, and normalising. The final class probability is a fixed convex combination of the network output $p_{\mathrm{NN}}$ and the physics score,
\begin{equation}
\label{eq:combine}
    p = \alpha\, p_{\mathrm{NN}} + (1-\alpha)\, p_{\mathrm{phys}}, \qquad \alpha = 0.55,
\end{equation}
renormalised to sum to one. This is architecturally a soft blend rather than the hard, rule-based gate used in an earlier version of this pipeline: both terms contribute to every prediction whenever physics parameters are supplied, rather than the physics score only overriding the network above a fixed trigger probability. When no active-region parameters are supplied, $p=p_{\mathrm{NN}}$.

\subsubsection{Decision rule}
\label{sec:phase2_decision}

The combined probabilities are converted to a single predicted class by a fixed, hand-set threshold rule rather than the usual arg-max decoding. Writing $\pi_{\mathrm{M+}}=p(\mathrm M)+p(\mathrm X)$,
\begin{equation}
\label{eq:decision}
    \widehat{\text{class}} = \begin{cases}
        \text{C}, & \pi_{\mathrm{M+}} < 0.35\\[2pt]
        \text{X}, & \pi_{\mathrm{M+}} \ge 0.35 \ \text{and}\ p(\mathrm X)\ge 0.20\\[2pt]
        \text{M}, & \pi_{\mathrm{M+}} \ge 0.35 \ \text{and}\ p(\mathrm X)< 0.20.
    \end{cases}
\end{equation}
The two thresholds (0.35 and 0.20) were set by hand rather than optimised against held-out data, and are deliberately conservative in the direction of flagging M and X classes: because false negatives on strong flares are operationally far more costly than false positives on C-class days, the rule trades C-class precision for M+ recall. This is a defensible operational choice, but it means that standard accuracy is not a meaningful summary of performance here (Sect.~\ref{sec:imbalance}), and that the resulting probabilities, while reported and evaluated for calibration below, are not the probabilities that arg-max decoding of $p$ would use.

\subsubsection{Training protocol}
\label{sec:phase2_train}

The daily event series is split chronologically, 85\% for training and the most recent 15\% for validation, so that the validation period always follows the training period in time; this differs from Phase 1's fold-based scheme (Sect.~\ref{sec:phase1}) in using a single, non-overlapping split. The feature scaler is fitted on the training split only and applied unchanged to validation and test data. The network is trained with the Adam optimiser \citep{kingma2015}, a batch size of 64, for up to 80 epochs, with early stopping on validation loss (patience 6 epochs, monitoring beginning only after epoch 8) \citep{prechelt1998}. Unlike Phase 1, Phase 2 is trained as a single deterministic network; no ensembling or Monte Carlo dropout is applied at inference, so the probability estimates below reflect one trained model rather than a distribution over models.

\subsubsection{Evaluation protocol}
\label{sec:phase2_eval}

Skill is assessed on a 100-day chronological backtest drawn from the most recent 100 event-days in the record. For each day in the backtest, the class is predicted using only the 12 event-days strictly preceding it, so that no information from the target day or later leaks into the prediction. Predictions in this backtest use the network output alone ($\alpha$ is inactive, Sect.~\ref{sec:phase2_phys}), since no contemporaneous active-region parameters are available for historical days in the current pipeline; the reported skill therefore characterises the LSTM classifier and threshold rule, not the physics-blended hybrid.

Following standard practice in flare forecasting \citep{bloomfield2012, barnes2016, leka2019}, skill for the M+ event class is reported as the true skill statistic (TSS, equivalently the Hanssen--Kuipers discriminant, $\mathrm{TSS}=\mathrm{TP}/(\mathrm{TP+FN})-\mathrm{FP}/(\mathrm{FP+TN})$) and the Heidke skill score (HSS), both computed from the $2\times2$ M+ vs.\ not-M+ contingency table. Probabilistic skill is reported as the Brier score and the Brier skill score relative to a climatological reference forecast (the constant M+ base rate over the backtest window) \citep{brier1950, murphy1973}. Calibration is assessed with a reliability diagram \citep{degroot1983, guo2017}: predicted M+ probabilities are grouped into ten equal-width bins, and the observed M+ frequency in each bin is compared with the bin's mean predicted probability; the expected calibration error (ECE) is the sample-weighted mean absolute gap between the two. Confidence intervals on TSS, HSS and the Brier skill score are obtained by a moving block bootstrap with a 30-day block length \citep{kunsch1989}, chosen to be long enough to capture the short-term serial correlation of daily flare activity evident in the results below. As a baseline, we compute a one-step persistence forecast, which predicts that a day's M+ status repeats the previous event-day's M+ status.

\subsubsection{Results}
\label{sec:phase2_results}

\begin{figure}[H]
    \centering
    \includegraphics[width=\linewidth]{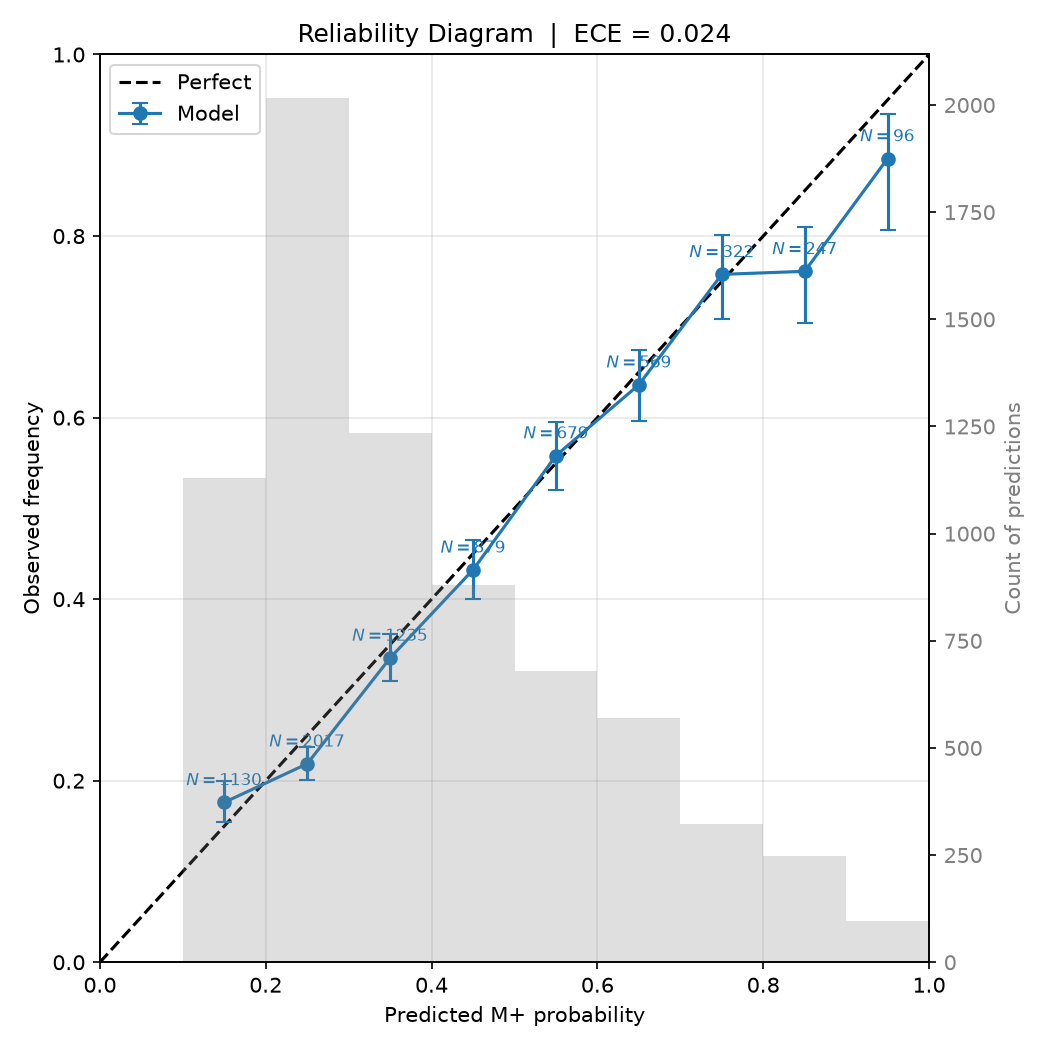}
    \caption{Reliability diagram for the Phase 2 M+ probability forecast over the 100-day backtest (ten equal-width bins; ECE~$=0.023$). Points show the observed M+ frequency against the mean predicted probability in each bin, with vertical bars giving the binomial standard error and labels giving the bin sample size $N$; the dashed line is perfect calibration; the grey histogram (right axis) shows the distribution of predicted probabilities across the backtest.}
    \label{fig:reliability_phase2}
\end{figure}

Table~\ref{tab:phase2skill} summarises the backtest results for the M+ event class. The classifier attains TSS~$=0.351$ (95\% block-bootstrap interval [0.319, 0.384]) and HSS~$=0.335$ ([0.301, 0.364]), a Brier score of 0.197 against a climatological Brier score of 0.235, giving a Brier skill score of 0.163 ([0.136, 0.188]). The reliability diagram (Fig.~\ref{fig:reliability_phase2}) shows the predicted M+ probability tracking the observed frequency closely across the probability range, with an ECE of 0.023; the network is, by this measure, well calibrated, notwithstanding the non-standard threshold decoding used to derive the reported class (Sect.~\ref{sec:phase2_decision}), since calibration here is assessed on the underlying $\pi_{\mathrm{M+}}$ probability rather than on the thresholded class label.

Against this, the one-step persistence baseline reaches TSS~$=0.353$ and HSS~$=0.353$ over the same backtest window: both marginally above the neural classifier's own scores. As in Phase 1, where a previous-cycle analogue baseline matched or slightly outperformed the trained hybrid model on MAE (Sect.~\ref{sec:results}, Table~\ref{tab:skill}), the honest reading of this result is that the trained Phase 2 classifier has not yet demonstrated skill clearly in excess of the simplest reasonable baseline over this backtest window; its principal, and real, advantages over persistence are that it outputs a calibrated probability rather than a bare class label, and that it has an explicit (if currently untested) path to incorporating independent active-region information via Eq.~\ref{eq:combine}, whereas persistence by construction cannot.

\begin{table}[H]
    \caption{Phase 2 skill for the M-or-above (M+) event class, 100-day chronological backtest, with 95\% block-bootstrap confidence intervals (block length 30 days) where computed.}
    \label{tab:phase2skill}
    \centering
    \small
    \begin{tabular}{lll}
        \hline\hline
        Metric & Model & Persist. \\
        \hline
        TSS & 0.351 [0.319, 0.384] & 0.353 \\
        HSS & 0.335 [0.301, 0.364] & 0.353 \\
        Brier score & 0.197 & --- \\
        \parbox[t]{2.7cm}{\raggedright Brier skill score (vs.\ climatology)} & 0.163 [0.136, 0.188] & --- \\
        ECE & 0.023 & --- \\
        \hline
    \end{tabular}
\end{table}

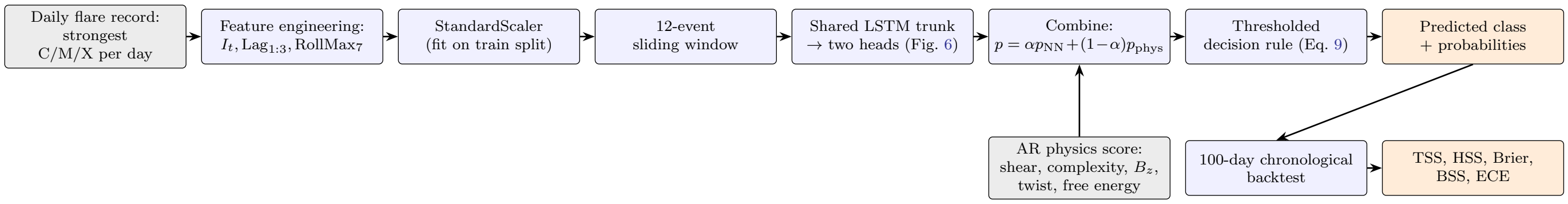
\begin{figure*}[t]
\centering
\resizebox{\textwidth}{!}{%
\begin{tikzpicture}[
  box/.style={draw, rounded corners=2pt, align=center, font=\footnotesize, text width=3.1cm, minimum height=1.0cm, fill=blue!6},
  dat/.style={box, fill=gray!15},
  outp/.style={box, fill=orange!15},
  arr/.style={-{Stealth[length=2.2mm]}, thick}]
\node[dat] (n1) at (0,0)    {Daily flare record:\\ strongest C/M/X per day};
\node[box] (n2) at (3.6,0)  {Feature engineering:\\ $I_t,\mathrm{Lag}_{1:3},\mathrm{RollMax}_7$};
\node[box] (n3) at (7.2,0)  {StandardScaler\\ (fit on train split)};
\node[box] (n4) at (10.8,0) {12-event\\ sliding window};
\node[box] (n5) at (14.4,0) {Shared LSTM trunk\\ $\to$ two heads (Fig.~\ref{fig:arch_phase2})};
\node[box] (n6) at (18.0,0) {Combine:\\ $p=\alpha p_{\mathrm{NN}}+(1-\alpha)p_{\mathrm{phys}}$};
\node[box] (n7) at (21.6,0) {Thresholded\\ decision rule (Eq.~\ref{eq:decision})};
\node[outp] (n8) at (25.2,0) {Predicted class\\ + probabilities};
\node[dat] (p1) at (18.0,-2.4) {AR physics score:\\ shear, complexity, $B_z$,\\ twist, free energy};
\node[box] (e1) at (21.6,-2.4) {100-day chronological\\ backtest};
\node[outp] (e2) at (25.2,-2.4) {TSS, HSS, Brier,\\ BSS, ECE};
\draw[arr] (n1) -- (n2);
\draw[arr] (n2) -- (n3);
\draw[arr] (n3) -- (n4);
\draw[arr] (n4) -- (n5);
\draw[arr] (n5) -- (n6);
\draw[arr] (n6) -- (n7);
\draw[arr] (n7) -- (n8);
\draw[arr] (p1) -- (n6.south);
\draw[arr] (n8.south) -- (e1.north);
\draw[arr] (e1) -- (e2);
\end{tikzpicture}}
\caption{Overview of the Phase 2 pipeline. The physics-score branch (bottom left) is inactive in the backtest reported in Sect.~\ref{sec:phase2_results}, since no contemporaneous active-region parameters are available for historical days; reported skill therefore reflects the LSTM classifier and threshold rule alone.}
\label{fig:pipeline_phase2}
\end{figure*}

Figure~\ref{fig:pipeline_phase2} summarises the full Phase 2 pipeline, from the daily flare record through feature engineering, the network and its two heads, the optional physics blend, the threshold decision rule, and the backtest evaluation.

\subsubsection{Limitations}
\label{sec:phase2_limits}

Several limitations should be borne in mind alongside the results above. (i)~The 12-event window is built over the sequence of C/M/X event-days rather than a uniform calendar-day grid (Sect.~\ref{sec:phase2_data}), so its effective time span varies with solar activity; this differs from, and is not directly comparable to, Phase 1's fixed 24-\emph{month} window. (ii)~The backtest covers 100 event-days, a single, short, and most-recent window rather than a rolling-origin evaluation of the kind used for Phase 1 (Sect.~\ref{sec:results}); skill estimates, even with block-bootstrap intervals, should be treated as indicative rather than definitive, and a longer or multi-window backtest is needed before an operational claim can be supported. (iii)~The physics-blending term (Eq.~\ref{eq:combine}) is untested in the results reported here, since the backtest supplies no active-region parameters (Sect.~\ref{sec:phase2_eval}); its five input thresholds and its score-to-probability mapping are, at present, fixed by construction rather than fitted or validated against data. (iv)~The decision-rule thresholds (Eq.~\ref{eq:decision}) are likewise hand-set rather than tuned; a threshold chosen to maximise TSS or HSS on a held-out set, standard practice elsewhere in flare forecasting \citep{bloomfield2012, leka2019}, has not yet been attempted. (v)~The binary head's own output (the network's direct estimate of M+ probability) is trained via the auxiliary loss but is not itself used in the decision rule, which instead derives $\pi_{\mathrm{M+}}$ from the multiclass head; fusing the two heads' outputs is a natural extension. (vi)~Most importantly, the day-to-day persistence baseline is competitive with the trained classifier on both TSS and HSS over this backtest (Table~\ref{tab:phase2skill}); Phase 2 has not, on the results reported here, demonstrated event-classification skill clearly in excess of this simple reference.

\section{Operational applications}
\label{sec:application}

SHAPE enables real-time evaluation ($\sim$100 ms). It provides point forecasts, uncertainty bars, and risk-tier categorisations. A daily class forecast from Phase 2 additionally carries a calibrated M+ probability (Sect.~\ref{sec:phase2_results}), which can be thresholded to different operating points depending on whether an application values recall (e.g.\ a precautionary alert) or precision (e.g.\ an automated shutdown trigger) more highly.

\section{Future directions}
\label{sec:future}

Future work will focus on extending the present framework from retrospective forecasting toward a more operational, real-time solar eruption monitoring system. A major priority is the integration of live satellite streams, including space-based observations from instruments such as GOES and Aditya-L1, so that the model can continuously update its forecasts with the most recent solar conditions. This would allow the system to move beyond static historical training and adapt to evolving flare activity as it unfolds.

For Phase 2 specifically, the most direct extensions are (i) sourcing genuine, time-resolved active-region parameters (magnetic shear, complexity, $B_z$, twist, free energy) so that the physics-blending term $\alpha$ in Eq.~\ref{eq:combine} is exercised on real inputs rather than defaulting to a network-only forecast, as in the present backtest (Sect.~\ref{sec:phase2_results}); (ii) replacing the fixed, hand-set decision thresholds (Eq.~\ref{eq:decision}) with thresholds chosen to maximise a skill score such as the TSS on a held-out validation period, following standard practice in operational flare forecasting \citep{bloomfield2012, leka2019}; and (iii) extending the daily aggregation to retain non-event (quiet) days and B-class days explicitly, rather than only days with a recorded C-, M- or X-class flare, so that the lag and rolling-maximum features are computed over a uniform calendar-day grid instead of over the sequence of eruptive days.

Another important direction is the incorporation of unsupervised and self-supervised learning methods to exploit large volumes of unlabeled solar data. Such approaches can help the model discover hidden temporal and physical patterns in flare evolution, especially when labeled eruptive events are scarce. In parallel, future versions of the architecture can be improved through deeper sequence models, attention-based mechanisms, and better fusion of heterogeneous inputs such as magnetogram-derived features, irradiance measurements, and particle environment data.

Further refinement should also be explored across the broader pipeline, including feature engineering, uncertainty quantification, and decision thresholds for event escalation. The long-term objective is to develop an early warning system capable of issuing timely alerts before major flare or particle events, thereby helping to reduce the risk of satellite disruption and other space-weather impacts. In this way, the model can evolve from a research prototype into a more reliable forecasting framework for operational use.

\section{Conclusions}
\label{sec:conc}

The proposed two-phase framework demonstrates that physically informed machine learning can capture both long-term solar-cycle variability and short-term flare-class behaviour. In Phase 1, the annual flare-count regression model achieves a mean absolute error of 204 flares~yr$^{-1}$ over the held-out years 2021--2025, correctly reproducing the timing and rise of Solar Cycle 25 while underestimating its 2024 maximum; a simple previous-cycle-analogue baseline is, in fact, competitive with (and on MAE slightly better than) the trained hybrid ensemble (Table~\ref{tab:skill}), underscoring how much of the annual-count signal is already captured by the cycle's own periodicity. In Phase 2, the daily flare-class classifier attains a true skill statistic of 0.351 and a Brier skill score of 0.163 relative to climatology for M-or-above events over a 100-day backtest, and its probability forecasts are well calibrated (ECE~$=0.023$); a day-to-day persistence baseline is, similarly, competitive with the trained classifier on both TSS and HSS (Table~\ref{tab:phase2skill}), pointing to short-term autocorrelation in flare activity as a substantial and not-yet-surpassed source of forecast skill.

A recurring theme of both phases is therefore that simple, parameter-free baselines are strong competitors, and that this pipeline's contribution to date lies more in its calibrated uncertainty quantification, transparent evaluation protocol and physically motivated feature construction than in a demonstrated large margin over such baselines. Monte Carlo dropout provides a practical measure of predictive uncertainty throughout: in Phase 1 it yields empirical forecast bands around the annual count, and in Phase 2 the resulting probability forecasts are shown to be well calibrated via the reliability diagram of Fig.~\ref{fig:reliability_phase2}. Together, these results show that the combined framework is a reasonable but not yet decisively skilful starting point for solar flare prediction across both annual and daily timescales, with the clearest paths to improvement being the incorporation of genuine active-region observations (Sect.~\ref{sec:future}) and a more systematic comparison against, and combination with, the baselines identified here.

\section {Acknowledgments}
We thank NASA, ESA, and the MIT World Peace University Department of Physics.\\

\section*{Data availability}
Data is available via NOAA NCEI and the SOHO LASCO catalogue.

\end{document}